\documentclass[prl,aps,twocolumn,showpacs,groupedaddress,superscriptaddress,amsmath,amssymb,10pt]{revtex4}
\usepackage{url,graphicx, color}
\usepackage{dcolumn}
\usepackage{bm}
\usepackage{epsfig,amsmath}
\usepackage{amssymb,bm}
\usepackage{color}

\def\vec#1{{\bf #1}}

\begin{document}

\title{Nanoscopic friction under electrochemical control}
\author{A.~S.~de Wijn}
\email{astrid@dewijn.eu}
\affiliation{Department of Physics, Stockholm University, 106 91 Stockholm, Sweden}
\author{A.~Fasolino}
\affiliation{Radboud University Nijmegen, Institute for Molecules and Materials, Heyendaalseweg 135, 6525AJ Nijmegen, the Netherlands} 
\author{A.~Filippov}
\affiliation{Donetsk Institute for Physics and Engineering of NASU, 83144, Donetsk, Ukraine}
\author{M.~Urbakh}
\affiliation{School of Chemistry, Tel Aviv University, 69978 Tel Aviv, Israel}

\begin{abstract}
  We propose a theoretical model of friction under electrochemical conditions focusing on the interaction of a force microscope tip with adsorbed polar molecules of which the orientation depends on the applied electric field.
  We demonstrate that the dependence of friction force on the electric field is determined by the interplay of two channels of energy dissipation: (i) the rotation of dipoles and (ii) slips of the tip over potential barriers.
  We suggest a promising strategy to achieve a strong dependence of nanoscopic friction on the external field based on the competition between long range electrostatic interactions and short range chemical interactions between tip and adsorbed polar molecules. 
\end{abstract}

\maketitle

Control of friction during sliding is extremely important for a large variety of applications~\cite{UrbakhNature2004,VanossiRevModPhys2013}.
A unique path to control and ultimately manipulate the forces between material surfaces is through an applied electric field.
By varying the applied potential,
the electrode surface can quickly and reversibly be modified either with adsorbed (sub)monolayer or multilayers, or via the oxidation and reduction of surfaces, or deposition of ultrathin films~\cite{SchmicklerSantos,Gileadi}.
Thus, friction force microscopy (FFM) measurements under electrochemical conditions~\cite{NielingerPhysChemChemPhys2007,HausenElectrochimActa2007,LabudaRevSciInstrum2010,HausenElectrochimActa2011,SweeneyPRL2012,ArgibayTribLett2012}
may offer significant advantages in comparison to those between dry surfaces.

Several experimental and theoretical studies of electrochemical interfaces demonstrated that
the orientation of polar molecules adsorbed at electrode surfaces is potential dependent. 
Water molecules at electrode/electrolyte interfaces reorient from ``oxygen-up'' to ``oxygen-down'' as the potential on the electrode changes from negative to positive~\cite{MottElectrochimActa1961,NagyJChemSocFaradayDiscuss1992,ToneyNature1994,AtakaJPhysChem1996}.
Another extensively studied system
 is pyridine adsorbed on gold electrodes~\cite{StolbergJElectroanalChem1987,CaiLangmuir1998}.
{Recent FFM measurements combined with cyclic voltammetry~\cite{Baltruschat63rd2012} have shown that friction depends strongly on the orientation of the molecules and is five times higher when the molecules are parallel to the substrate compared to their vertical orientation.  The molecule orientation can be changed either by changing their concentration or by an external field.  In the latter case, the friction shows an intense peak around values of the field where the change of orientation takes place.}
Recent FFM measurements in UHV have also shown strong sensitivity of nanoscopic friction to the orientation of surface molecules~\cite{FesslerApplPhysLett2011}.
Investigating the impact of potential-dependent orientation of adsorbed molecules on friction offers a new perspective on active control of friction forces
through reversible molecular reorientation.

In spite of the first successful experimental studies of nanoscopic friction under potential control~\cite{NielingerPhysChemChemPhys2007,HausenElectrochimActa2007,LabudaRevSciInstrum2010,HausenElectrochimActa2011,SweeneyPRL2012,ArgibayTribLett2012}, so far there have been no theoretical or numerical studies of the effect of electric fields on friction.
We do not know what the mechanism is behind the observed variation of friction with electrostatic potential, nor in which systems significant reversible variation of the friction can be achieved.

In this Letter we propose a minimal model for the description of the effect of potential dependent reorientation of adsorbed polar molecules on nanoscopic friction.
We investigate the effect of an applied potential and of tip-molecule interactions on energy dissipation and discuss conditions for controlled variation of the friction.

\begin{figure}
\epsfig{figure=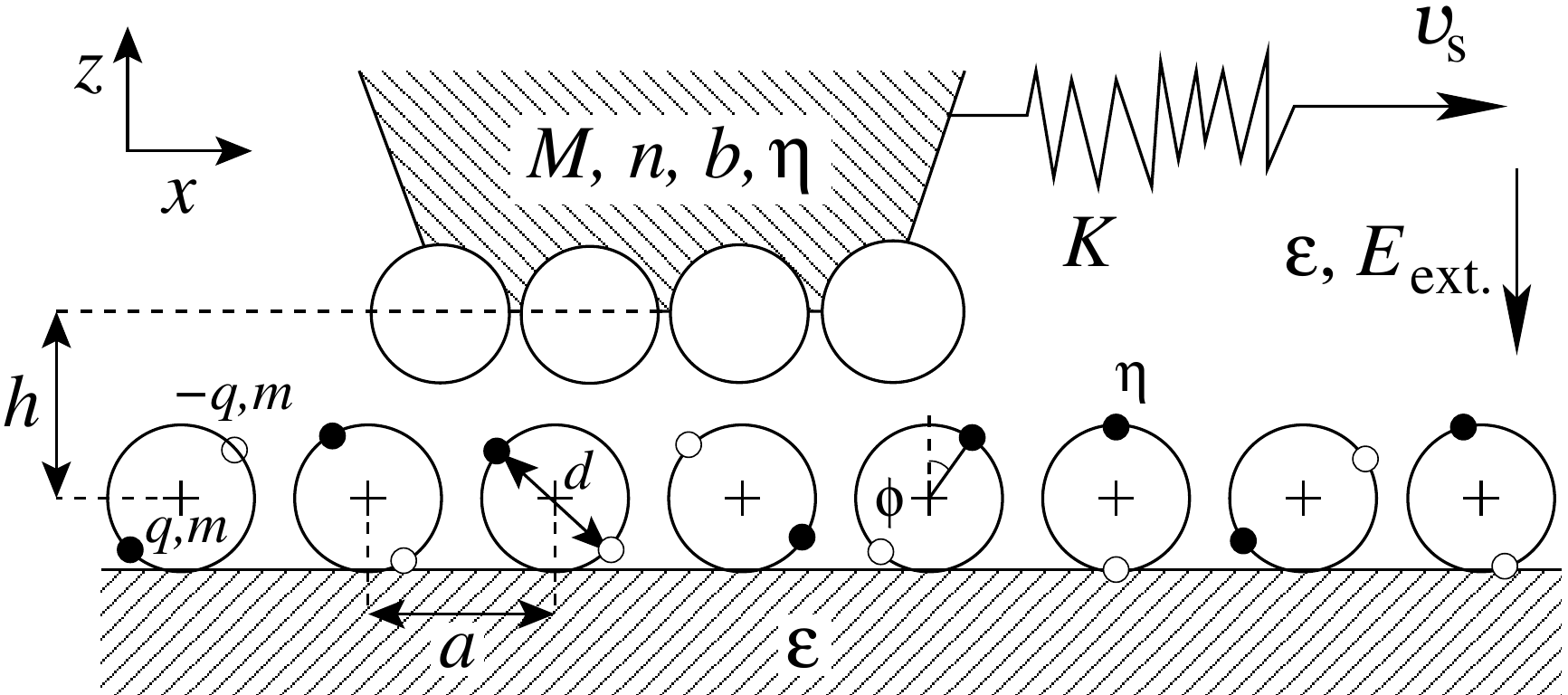,width=8.2cm}~~~
\caption{
Sketch of the model.
A tip is dragged over a surface covered by a monolayer of polar molecules, in the presence of an external electric field  $E_\mathrm{ext}$.
\label{fig:model}
}
\end{figure}

Our model is illustrated in Fig.~\ref{fig:model}. 
To mimic an FFM experiment, we consider a tip with mass $M$ and center-of-mass coordinate $X$, coupled by a spring of spring constant $K$ to a support that moves at constant velocity $v_\mathrm{s}$. Polar molecules adsorbed at the electrode are represented by rigid interacting dipoles with fixed centers of mass which, in the absence of external interactions, lie horizontally, head-to-tail on the surface at fixed positions. 
The external electric field $E_\mathrm{ext}$ is directed perpendicular to the surface in the $z$ direction. Both electrostatic and short range chemical interactions between the dipoles and the tip are considered. The tip is dragged along the surface at a height $h$ from the centers of the dipoles and 
the time average of the lateral force $F_\mathrm{lat}$, i.e.\ the tension in the spring, gives the friction force.

The simulations are performed with parameters describing water molecules at the Pt surface, {which is a commonly-used prototype system in simulations~\cite{NagyJChemSocFaradayDiscuss1992},} but 
the same approach applies also to other, larger, adsorbed polar molecules.

Each dipole consists of two charges $+q$ and $-q$ and masses $m$ separated by a fixed distance $d$. The $N$  dipoles  are arranged in a chain with spacing $a$  with periodic boundary conditions and can only rotate around their center of mass.
We choose $q=6.44\times 10^{-20}$C, $m=6.42\times 10^{-26}$kg and $d=0.958$\AA{} to have the dipole moment and moment of inertia of water molecules,  $a=2.77$\AA{} as the lattice spacing of Pt and  $h=d$. 
{While rotation around the center of mass is justified for water molecules adsorbed on metal electrodes~\cite{MottElectrochimActa1961,NagyJChemSocFaradayDiscuss1992,ToneyNature1994,AtakaJPhysChem1996}, it is also possible to anchor the dipoles at one end.  This may be more realistic for some adsorbates.}
Unless otherwise stated $N=128$.
The tip is modelled as an array of $n$ atoms ($n=1, 2, 4, 8, 16$) with interatomic distance $b$.  
The support velocity is $v_\mathrm{s}=10$m/s, which is sufficiently low for the system to exhibit stick-slip behavior (see supplementary material~\cite{supplementary}).

The dipoles interact with the charge of the tip and with their nearest neighbors through electrostatic interactions
$V_e({r})={q_1 q_2}/({4\pi\epsilon_0\epsilon {r}})$,
where ${r}$ is the distance between the two charges $q_1$ and $q_2$, and $\epsilon$ is the relative electric permittivity of the aqueous solution at the surface.
We use $\epsilon=5$, as estimated by measurements of the double layer capacitance at metal-electrolyte interfaces~\cite{SchmicklerSantos,KornyshevSolvation1986}.
The interaction between the tip and dipole charges is cut off at $Na/2$.
We neglect interactions between dipoles beyond nearest neighbors by assuming screening due to the ions located in the diffuse part of the double layer.

There may also be a short-range chemical interaction between the atoms of the tip and one or both ends of the dipoles.
This chemical interaction is modeled by the repulsive potential
\begin{align}
  V_\mathrm{c} ({r}) & = V_{\mathrm{c} 0} \exp(- {r}^2/\sigma_0^2)~,
\end{align}
where the energy and length scales, $V_{\mathrm{c} 0}$ and $\sigma_0$, are taken as $0.5$eV and $\frac12 a$ respectively.

The temperature $T$ is controlled by a Langevin thermostat with damping constant $\eta=1/$ps on the tip and dipoles.
The equations of motion thus take the form
\begin{align}
  \label{eq:eom1}
  \ddot{X} & = - \frac{K}{M} (X-v_\mathrm{s} t)
  - \frac{1}{M} \frac{\partial}{\partial X} V^{\mathrm{t-d}}
  \nonumber\\& \phantom{=} \strut
  - \eta \dot{X} + \xi_\mathrm{tip}(t)~,\displaybreak[0]\\
  \label{eq:eom2}
  \ddot{\phi}_i & = - \frac{1}{I} \frac{\partial}{\partial \phi_i}\left[ V^{\mathrm{d-d}}+V^{\mathrm{t-d}}\right]
  \nonumber\\ &\phantom{=}\strut
  - \eta {\dot\phi}_i + \xi_i(t)- E_\mathrm{ext} \sin\phi~, 
\end{align}
where $\phi_i$ is the angle of the $i$-th dipole with respect to the $z$-axis, $I$ is the moment of inertia of a dipole, and the potentials $V^{\mathrm{t-d}}$ and $V^{\mathrm{d-d}}$ describe the tip-dipole and dipole-dipole interactions
(see supplementary material~\cite{supplementary}), while $\xi_\mathrm{tip}(t)$ and $\xi_i(t)$ are the random forces of the thermostat.

\begin{figure}
\epsfig{figure=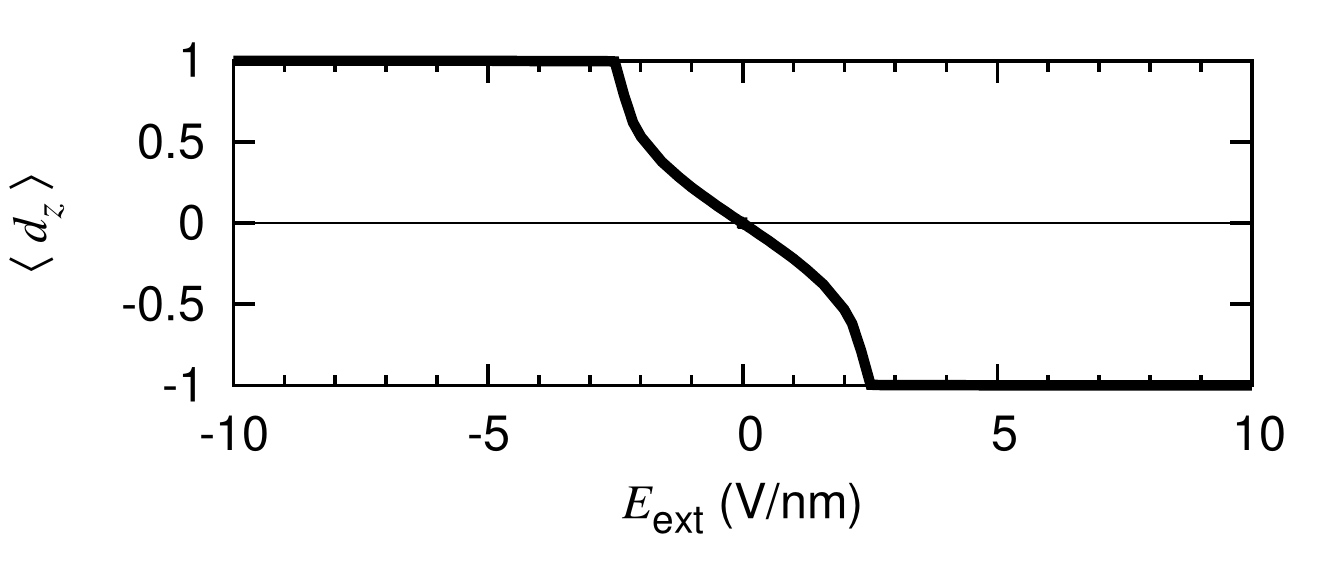,angle=270,width=8.4cm}
\vskip-\bigskipamount
\caption{
Dependence of the average $z$-component of the unit vector along the dipole direction, $d_z = \cos\phi$, on the electric field for a system of 512 dipoles equilibrated at $T=2.93$K without the tip.
\label{fig:op}
}
\end{figure}

In Fig.~\ref{fig:op} we show how the equilibrium configuration of the dipoles depends on the field $E_\mathrm{ext}$, in the absence of the tip.
A similar dipole model has been shown
to provide a good description of potential-dependent reorientation of adsorbed molecules studied by spectroscopic and capacitance measurements~\cite{SchmicklerSantos,ToneyNature1994,AtakaJPhysChem1996,HamelinElectroanalChem1996,MonroeJElectrochemSoc2009}.
There is an abrupt change in the equilibrium dipole orientation as a function of the field.
The molecules lay flat ($\langle d_z \rangle$=0, with ${d_z=\cos\phi}$ the $z$ component of the dipole unit vector) for $E_\mathrm{ext}=0$ and stand vertically ($\langle d_z\rangle=\pm 1$) for strong fields.
The transition occurs at field strength $E_\mathrm{ext}=E_\mathrm{ext}^\mathrm{trans}\approx
\pm 2.4$V/nm in our case, corresponding to the maxima in the variance of $\phi_i$.

We consider two limiting cases for the tip-dipole interaction: an electrostatic interaction between the charged tip and dipoles, and a short-range chemical interaction between the tip and one side of the dipole molecule. 

We now consider the channels through which energy is dissipated.
Energy is pumped into the system through the support, with average power $v_\mathrm{s} \langle F_\mathrm{lat} \rangle$ and through thermal noise at a rate of $\eta k T$ per degree of freedom.
Energy is removed from the system through the viscous damping of the dipoles and tip.
The power dissipated by the $i$-th dipole can be written as
$\dot\phi_i \cdot (\eta I \dot\phi_i) = 2\eta K^{\mathrm{d}}_{i}$,
where $K^{\mathrm{d}}_{i}$ is the kinetic energy of the dipole.
The contribution to the average lateral force due to dipole rotations away from equilibrium is thus
\begin{align}
  \label{eq:dissipation}
  F_{\mathrm{rot}} =  \frac{\eta}{v_\mathrm{s}}\sum_i (2 K^{\mathrm{d}}_{i} - k T)~.
\end{align}

\begin{figure}
\epsfig{figure=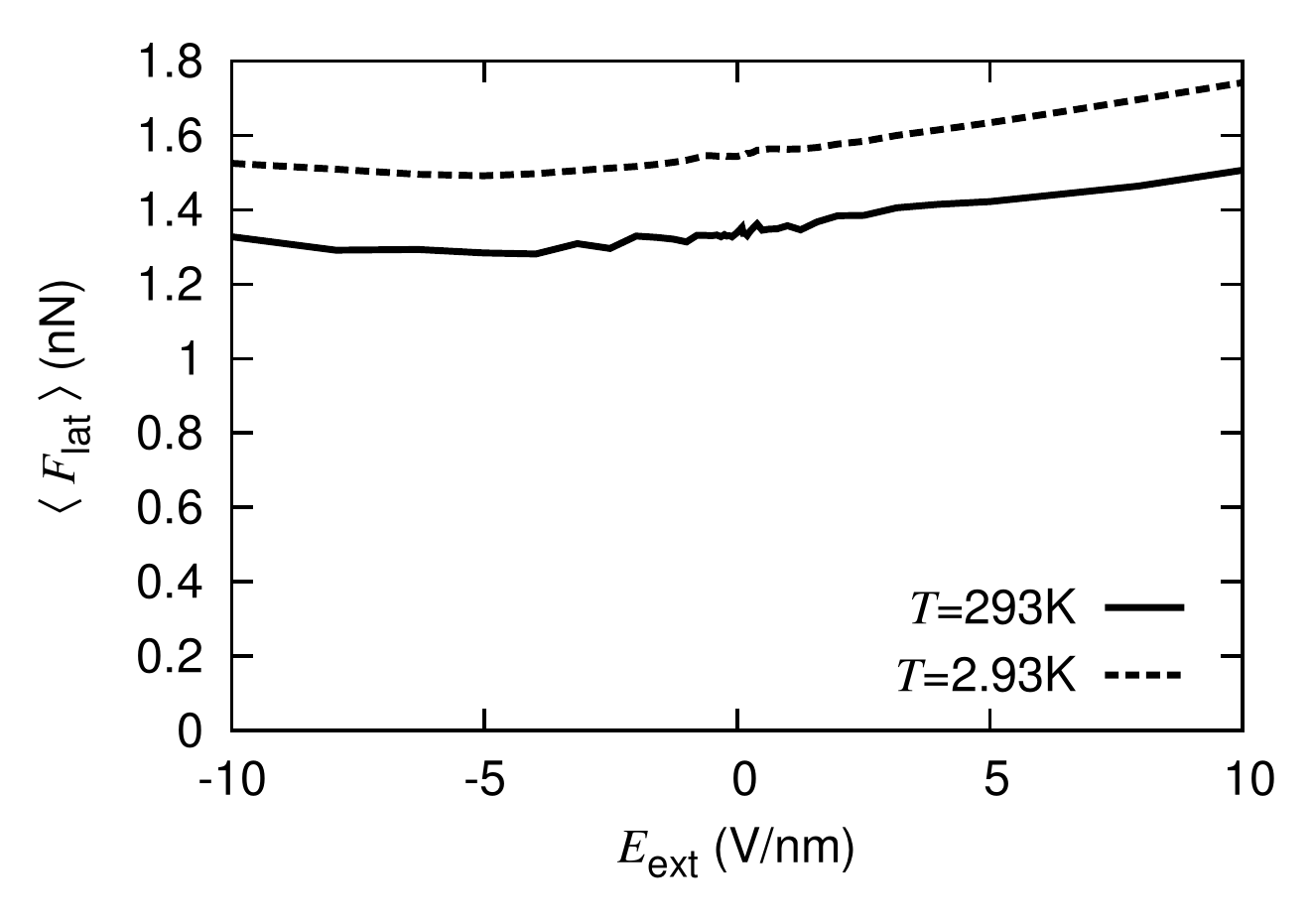,angle=270,width=8.4cm}
\vskip-4.35cm\hskip-0.9cm\epsfig{figure=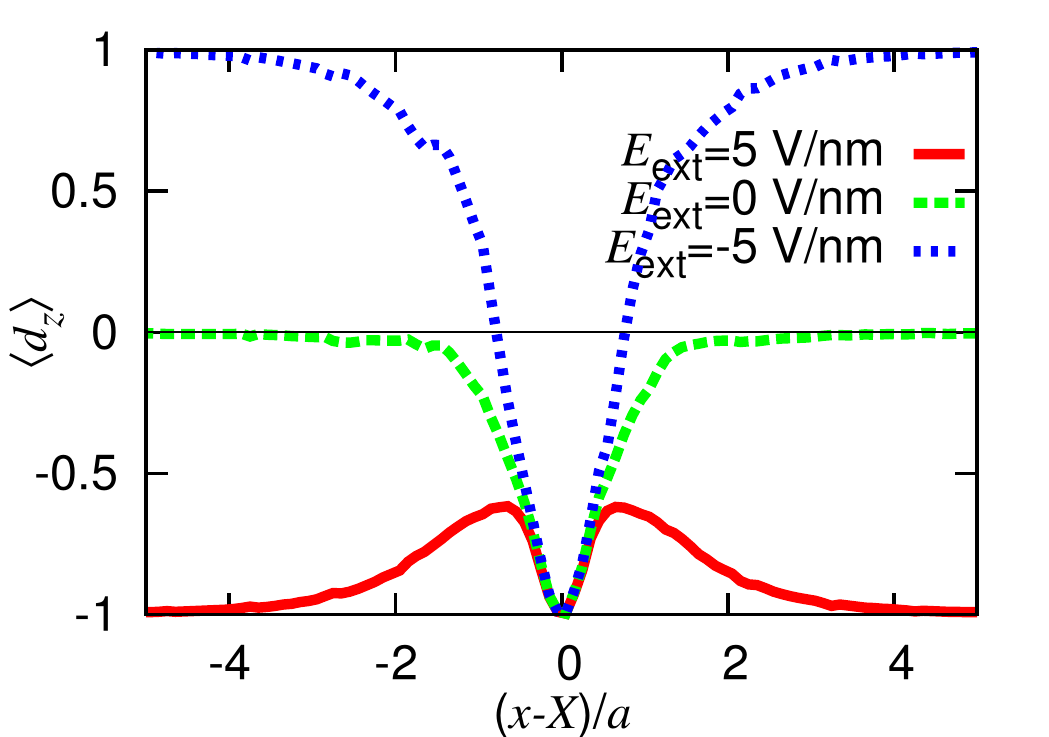,angle=270,width=4.3cm}
\vskip1.35cm
\vskip-\bigskipamount
\caption{
(color online)
Friction characteristics for a system with purely electrostatic interactions:
the friction force as a function of the field strength 
for two different temperatures, and
(inset) the average dipole orientation as a function of the distance of the dipoles from  the tip for three different fields strengths at $T=2.93$K.
\label{fig:electrostatic}
}
\end{figure}

In Fig.~\ref{fig:electrostatic} we show the dependence of the friction on the external field 
and the average dipole orientation as a function of the distance to the tip for purely electrostatic tip-dipole interactions.
Here we consider a point-like tip with a charge $Q=-2e$ that mimics a uniformly charged spherical tip. We see that the friction force has a very weak dependence on $E_\mathrm{ext}$.
For strong electrostatic tip-dipole interaction [$Qqd/(4 \pi \epsilon_0\epsilon h^2) > E_\mathrm{ext}qd$], which is the case for the entire range of fields shown in Fig.~\ref{fig:electrostatic}, the dipole orientation near the tip is dominated by the tip charge (see inset in Fig.~\ref{fig:electrostatic}), and it is independent of the applied field.
For negative fields the slip of the tip is accompanied by the reorientation of dipoles located below the tip
that can be considered as a molecular ball-bearing effect~\cite{RavivScience2002,BriscoeNature2006}.
In our simulations the tip-induced rotation of dipoles leads to a 30-40\% reduction of friction compared to the case of ``frozen'' dipoles.

We are forced to conclude that, in the presence of strong electrostatic interactions,
an external field cannot be used to affect the friction strongly.
The potential barrier that the tip must overcome to slip dominates the friction.
The contribution to the friction from the dipole rotations {in Fig.~\ref{fig:electrostatic}} is only about one percent, 0.1nN.
The height of the barrier is determined by the local configuration of dipoles in the vicinity of the tip.
Thus, the field has little effect on the friction force.
We note that, even if the dipoles near the tip could be reoriented, the long-range nature of electrostatic interactions imples that the coupling to the tip would not change drastically.

In the case of weak electrostatic tip-dipole interactions [$Qqd/(4 \pi \epsilon_0 \epsilon h^2) < E_\mathrm{ext}qd$] the energy dissipation through the dipole rotations can be a dominant contribution to the friction, and as a result the friction can strongly depend on the electric field. However, weak electrostatic interactions produce much less friction than is typically measured in experiments.  We have confirmed this in simulations with $h = 7d$.

\begin{figure}
\epsfig{figure=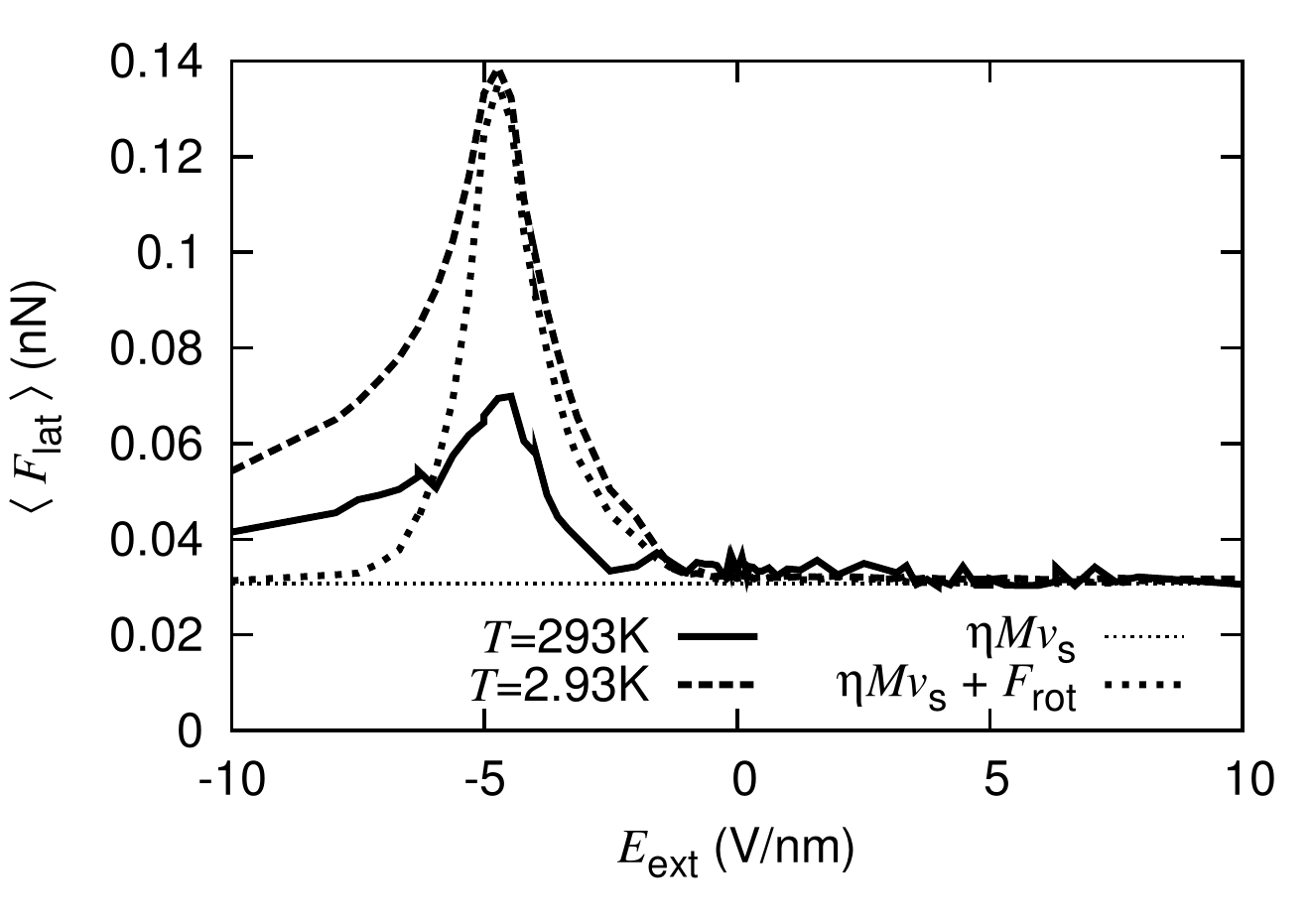,angle=270,width=8.4cm}
\vskip-5.5cm\hskip3.3cm\epsfig{figure=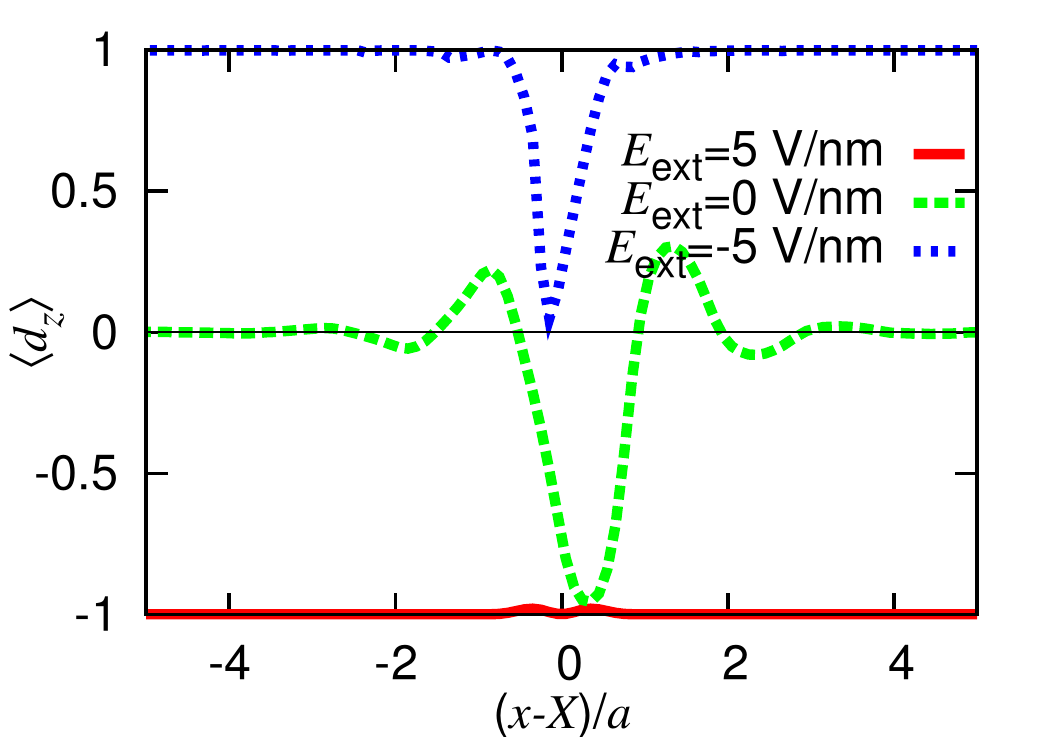,angle=270,width=4.3cm}
\vskip2.0cm
\caption{
(color online)
Friction for a system with a point-like tip $n=1$, with charge $Q=0$, but with chemical interactions between the tip and the negative end of the dipoles, plotted as a function of the field for two temperatures.  The contributions to the total dissipation from viscous friction
and dipole rotation at $T=2.93$~K is represented by the dashed line.
The peak of the friction force at $-5$V/nm is due to dissipation from dipole rotation.
The inset shows the average dipole orientation as a function of the distance to the tip for three different fields strengths at $T=2.93$K.
Additional material, including a movie, can be found in the supplementary material~\cite{supplementary}.
\label{fig:chemical}
}
\end{figure}

The scenario changes drastically if we add short-range chemical interactions.
We first consider a point-like tip with only chemical interactions with one (negative) side of the dipole molecule, i.e. $n=1, Q=0$.
In Fig.~\ref{fig:chemical} we show the friction force, which displays a strong peak around $E_\mathrm{ext}\approx-5$V/nm.
A temperature increase only reduces this effect without qualitative changes.
By comparing the calculated friction force with the dissipation by the dipoles [see Eq.~(\ref{eq:dissipation})] we see that the peak is largely due to the dipole rotations.

{This behavior is qualitatively similar to that observed for pyridine on gold in Ref.~\cite{Baltruschat63rd2012}, where a peak of the friction force appears to be related to the reorientation of the molecules.}
{The position of the peak and whether the friction is lower or higher for positive or negative values of the field depend on the details of the tip-molecule interaction and location of the site of strong chemical interaction.
}
{Our model shows higher friction when the dipoles are oriented with their chemically strongly interacting site pointed upwards (to the tip).  In the case of pyridine molecules, the most strongly interacting site is likely the side of the benzene ring or the N atom, not the terminating H atoms.  Consequently, friction is strongest when
the molecules lie flat on the surface.
We have also performed simulations for the case where both ends of the dipole interact chemically with the tip, and found two peaks, at positive and negative fields.
}

The mechanism of dissipation is very different than in the case of purely electrostatic interactions.
In the inset of Fig.~\ref{fig:chemical} we show the orientation of the dipoles as a function of the distance to the tip.
Contrary to the inset in Fig.~\ref{fig:electrostatic}, the orientation close to the tip depends on the field, so that the tip-dipole interaction does not wash out the effect of the field.
In the case of relatively weak chemical interactions the tip induces the reorientation of dipoles only in a limited range of fields [$V_{\mathrm{c} 0}\exp(-h^2/\sigma^2)<E_\mathrm{ext}qd<0$] that for our choice of parameters corresponds to -6V/nm$<E_\mathrm{ext}<0$.
In this interval of fields the dynamics in the dipole chain become the dominant mode of dissipation, and orientation waves can be created.
Outside the above interval of fields the friction force is determined by the energy dissipated by the tip during its slip over the barriers corresponding to a frozen (tip-independent) dipole configuration.
These barriers are very low for positive fields, where the negative (interacting) side of the dipole is far from the tip, and they are relatively high for $E<-6$V/nm, where the negative side of the dipole faces the tip.

\begin{figure}
\vskip0.7\bigskipamount
(a)\phantom{xxxxxxxxxxxxxxxxxxxxxxxxxxxxxxxxxxxxxxxxxxxx}
\vskip-2.2\bigskipamount
\epsfig{figure=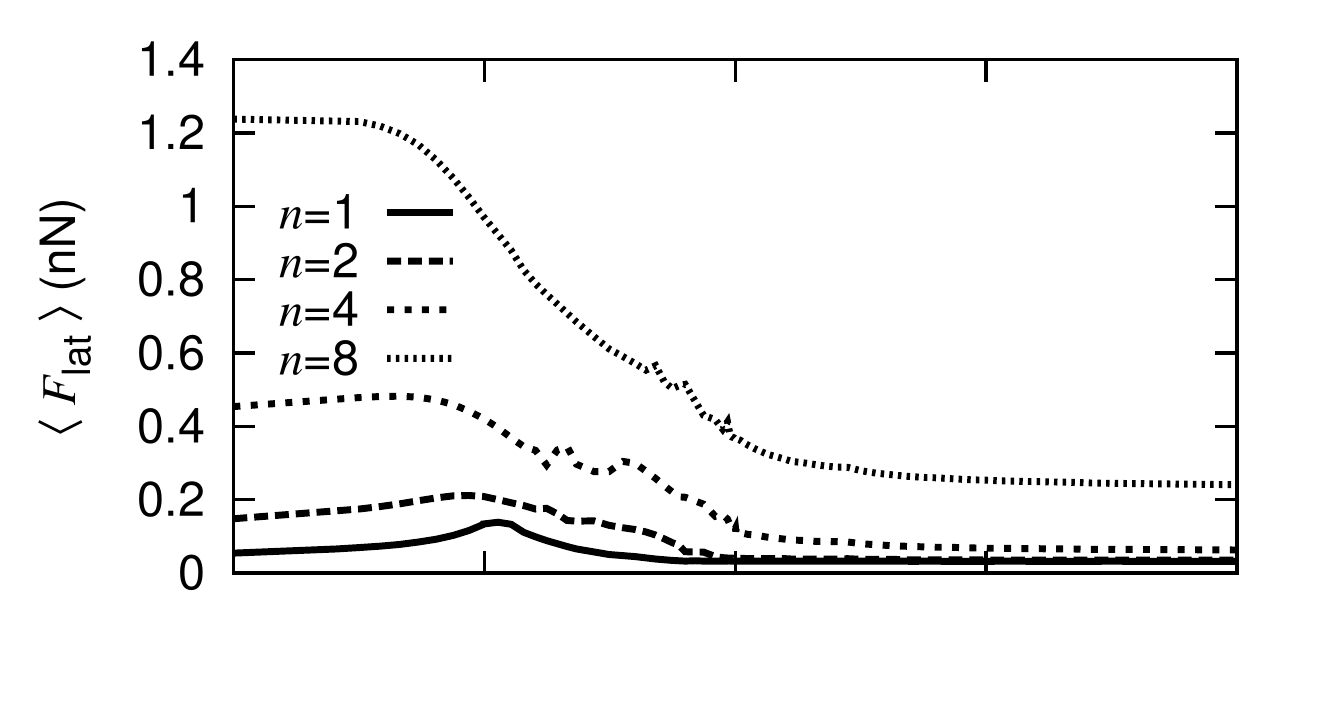,angle=270,width=8.4cm}
\vskip-4.2cm\hskip3.1cm\epsfig{figure=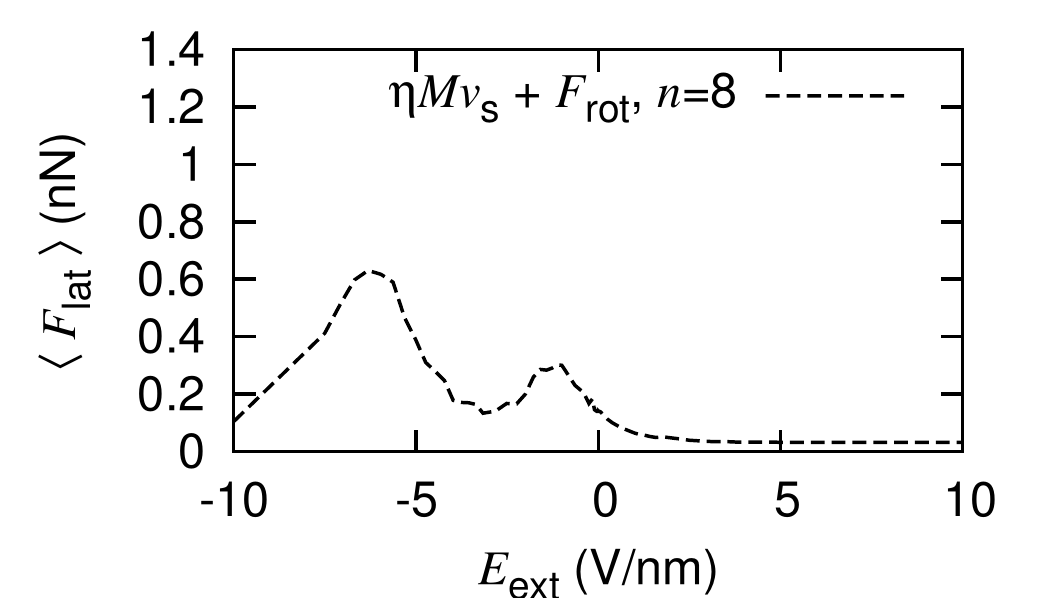, angle=270,width=4.3cm}
\vskip0.5cm
\vskip0.7\bigskipamount
(b)\phantom{xxxxxxxxxxxxxxxxxxxxxxxxxxxxxxxxxxxxxxxxxxxx}
\vskip-2.2\bigskipamount
\epsfig{figure=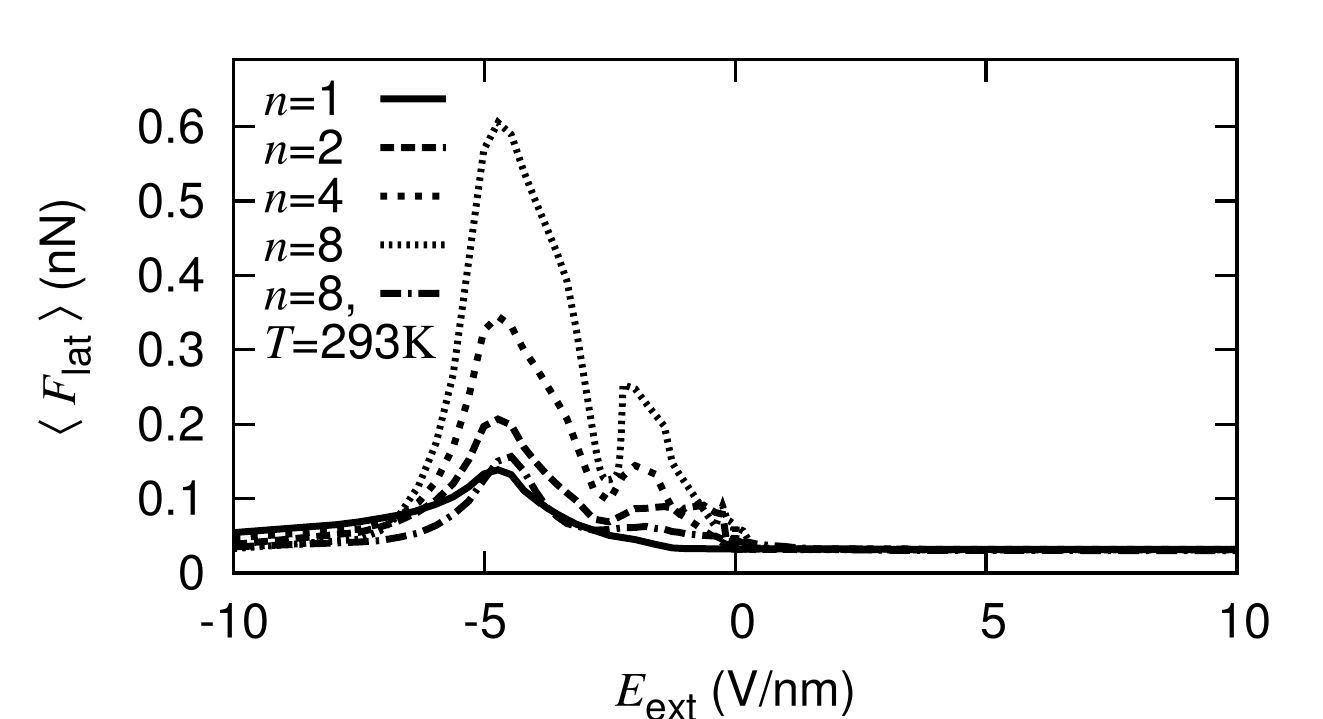,angle=270,width=8.4cm}
\vskip-4.2cm\hskip3.1cm\epsfig{figure=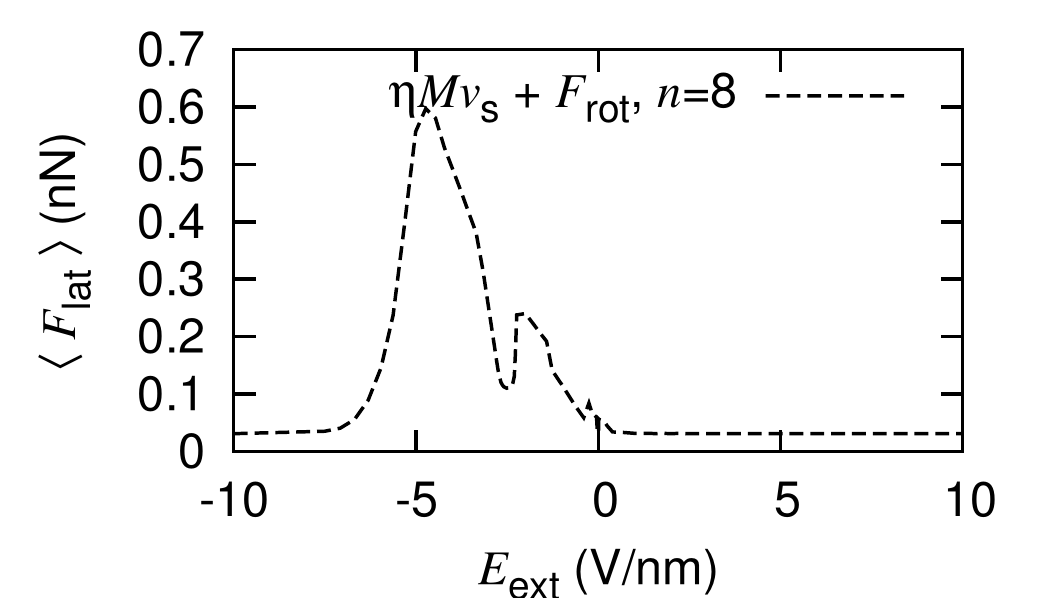, angle=270,width=4.3cm}
\vskip1.6cm
\caption{
The friction force as a function of the external field for extended commensurate tips of  different sizes and with lattice spacing commensurate $b=a$ (a) and incommensurate $b=\lambda a$ (b) with the dipole array.
The insets show the contributions from dipole rotation for the case of $n=8$.
For larger commensurate tips the high barrier dominates the friction force, while for incommensurate tips the dipole rotation dominates the friction force.
\label{fig:largetip}
}
\end{figure}

Although  the friction of the point-like tip with chemical interactions is strongly affected by the field, it is an order of magnitude smaller than what is typically measured in experiments.
In reality, a tip-surface contact is not point-like, but it has a size of the order of few nm$^2$, and it interacts simultaneously with a number of dipoles.  This leads to a corresponding increase of friction force.
We therefore now consider the effect of an extended tip.

In Fig.~\ref{fig:largetip} we show the friction force as a function of the external field for several tip sizes, for both commensurate ($b=a$) and incommensurate ($b=\lambda a$, with $\lambda\approx 1.618$ the golden ratio) contacts between the tip and the dipole array.
In Fig.~\ref{fig:largetip}a, we see that, for a commensurate tip, the dependence of the friction on the electric field changes with tip size and extends over a wider range of fields, while the typical value of the friction scales roughly linearly with the tip size.
For larger commensurate tips, the energy barrier that the tip must overcome to slip becomes larger and this starts to dominate the friction force.
Fig.~\ref{fig:largetip}b shows results for a tip with a lattice parameter incommensurate with respect to the dipole array.
In this case the effective potential corrugation does not grow with the tip size and the dissipation remains dominated by dipole rotation.
The
friction increases with the tip size.

Very close to the transition at $E_\mathrm{ext}^\mathrm{trans}$, there is a strong reduction in the friction.
This dip can be understood by considering that
the coupling between deviations of orientations of neighboring dipoles from the equilibrium ones changes sign near the transition, going through 0.
As a result, to leading order, energy cannot be transferred between dipoles, and can only be dissipated locally.
Movies of this effect can be found in the supplementary material~\cite{supplementary}.
{At higher temperatures, this effect is washed out, with the dip eventally disappearing, as can be seen also from the high-temperature simulations shown in Fig.~\ref{fig:largetip}b.  Such a dip is also not observed in the experiments of Ref.~\cite{Baltruschat63rd2012}.}

To summarize, a minimal model for the description of nanoscopic friction under electrochemical control has been proposed and investigated. We have considered two limiting types of interaction between the tip and molecules adsorbed at electrodes: (i) an electrostatic interaction between the charged tip and dipoles, and (ii) a short-range chemical interaction between the tip surface and one side of the dipole molecule.
We have demonstrated that the dependence of friction force on the electric field is determined by the interplay of two channels of energy dissipation: the rotation of dipoles and slips of the tip over potential barriers.
Enhancement of friction with accessibility of rotation degrees of freedom has been previously found in simulations of friction of single pyrrole molecules during diffusion on Cu(111)~\cite{cppyrrolethiophene}.
Here, we have found that friction can be efficiently tuned by an external electric field when the strength of the tip-dipole interaction is comparable with the dipole-electric field interactions.

\section{Acknowledgements}
  ASdW is supported by an Unga Forskare grant from the Swedish Research Council. MU acknowledges support by the Israel Science Foundation Grant and by the German-Israeli Project Cooperation Program (DIP).
  AF acknowledges support from the Foundation for Fundamental Research on Matter (FOM), which is part of the Netherlands Organisation for Scientific Research (NWO).
  Computational resources were provided by SNIC through Uppsala Multidisciplinary Center for Advanced Computational Science (UPPMAX). 
  ASdW is grateful to Roland Bennewitz and Lars Pettersson for discussions.
  ASdW and AEF acknowledge the hospitality of the School of Chemistry, Tel Aviv University.

  \clearpage
  
  \setcounter{page}{1}
  
  \setcounter{figure}{0}
  \renewcommand{\thefigure}{S\arabic{figure}}
  
  \setcounter{table}{0}
  \renewcommand{\thetable}{S\Roman{table}}
  
  \setcounter{equation}{0}
  \renewcommand{\theequation}{S\arabic{equation}}

  \begin{widetext}
  \begin{center}
  {\large \bf Nanoscopic friction under electrochemical control: supplementary material}
  \end{center}
  \bigskip
  \bigskip
  \end{widetext}
  
  \section{Potential energy functions}
  The potentials, $V^\mathrm{t-d}$ and $V^\mathrm{d-d}$, describing the tip-dipole and dipole dipole interactions respectively, can be written as a sum over positions of the tip atoms, $\vec{R}^\mathrm{t}_j$ (defined below), and the charges at the dipoles: 
  \begin{align}
    V^{\mathrm{t-d}} & = \sum_{i=0}^{N-1}\sum_{j=0}^{n-1} V^{\mathrm{tip}}_{i}(X_j)~,\\
    V^{\mathrm{d-d}} & = \sum_i V_\mathrm{dipole}(\phi_i,\phi_{i+1})~,
  \end{align}
  Here,
  \begin{align}
      V^{\mathrm{tip}}_{i}(X) & = \frac{qQ}{n} \sum_j [V_e(|\vec{R}^\mathrm{t}_j-\vec{r}_{i+}|)-V_e(|\vec{R}^\mathrm{t}_j-\vec{r}_{i-}|) ]
      \nonumber\\ &\phantom{=}\strut
      + \sum_j V_\mathrm{c}(|\vec{R}^\mathrm{t}_j-\vec{r}_{i+}|)
      ~,\\
      V_\mathrm{dipole}(\phi_i,\phi_j) & = qq [V_e(|\vec{r}_{i+}-\vec{r}_{j+}|) +V_e(|\vec{r}_{i-}-\vec{r}_{j-}|)
      \nonumber\\ &\phantom{=}\strut
      -V_e(|\vec{r}_{i+}-\vec{r}_{j-}|) -V_e(|\vec{r}_{i-}-\vec{r}_{j+}|)]~,
  \end{align}
  are the potential energies of interaction between the tip and dipole $i$ and between dipoles $i$ and $j$ respectively,
  while the positions of the positive and negative charges in the dipoles are given by $\vec{r}_{i+}$ and $\vec{r}_{i-}$, respectively.
  The potential energy functions and positions of the atoms are given by
  \begin{align}
      V_e({r}) & = \frac{1}{4 \pi \epsilon_0\epsilon {r}}~,\\
      V_\mathrm{c} ({r}) & = V_{\mathrm{c} 0} \exp(- {r}^2/\sigma_0^2)~,\label{eq:chemical}\\
      \vec{R}^\mathrm{t}_j & = (X_j,h)~,\\
      X_j & = X+b\left(j-\frac12(n-1)\right)~,\\
      \vec{r}_{i\pm} & = \left(ai\pm \frac12 d \sin\phi_i,\pm \frac12 d \cos\phi_i\right)~.
  \end{align}
  The equations of motion are integrated numerically using a fourth-order Runge-Kutta algorithm.

  \section{additional plots and video material}
  
  Here we include some additional plots and video material to further elaborate on the effects shown in Figs.~\ref{fig:chemical} and~\ref{fig:largetip} of the main article.

  \begin{figure}
  \vskip0.0\bigskipamount
  (a)\phantom{xxxxxxxxxxxxxxxxxxxxxxxxxxxxxxxxxxxxxxxxxxxx}
  \vskip-1.5\bigskipamount
  \vskip6.5\bigskipamount
  (b)\phantom{xxxxxxxxxxxxxxxxxxxxxxxxxxxxxxxxxxxxxxxxxxxx}
  \vskip-7.5\bigskipamount
  \epsfig{figure=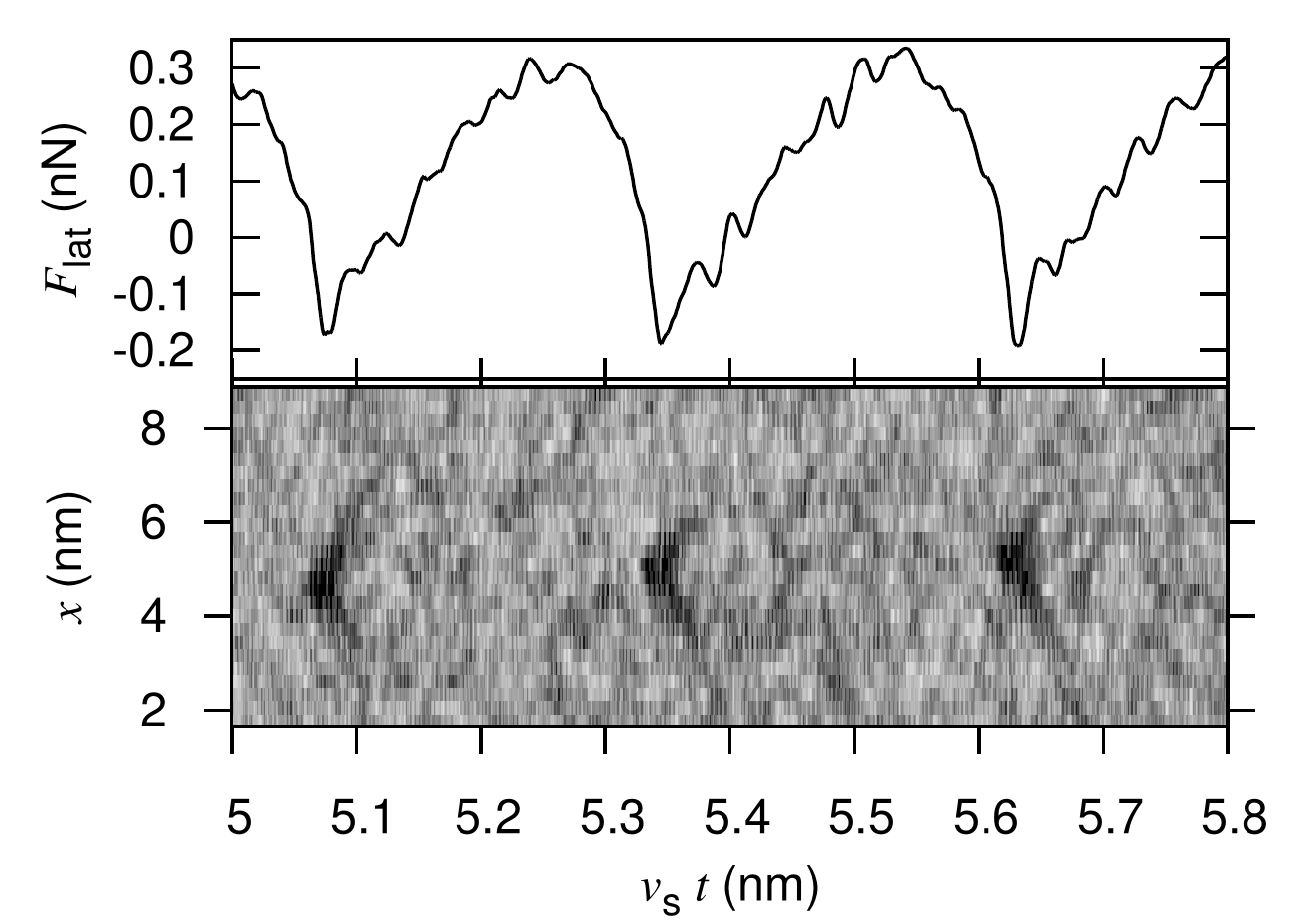,angle=270,width=8.4cm}
  \vskip0.0\bigskipamount
  (c)\phantom{xxxxxxxxxxxxxxxxxxxxxxxxxxxxxxxxxxxxxxxxxxxx}
  \vskip-1.5\bigskipamount
  \vskip6.5\bigskipamount
  (d)\phantom{xxxxxxxxxxxxxxxxxxxxxxxxxxxxxxxxxxxxxxxxxxxx}
  \vskip-7.5\bigskipamount
  \epsfig{figure=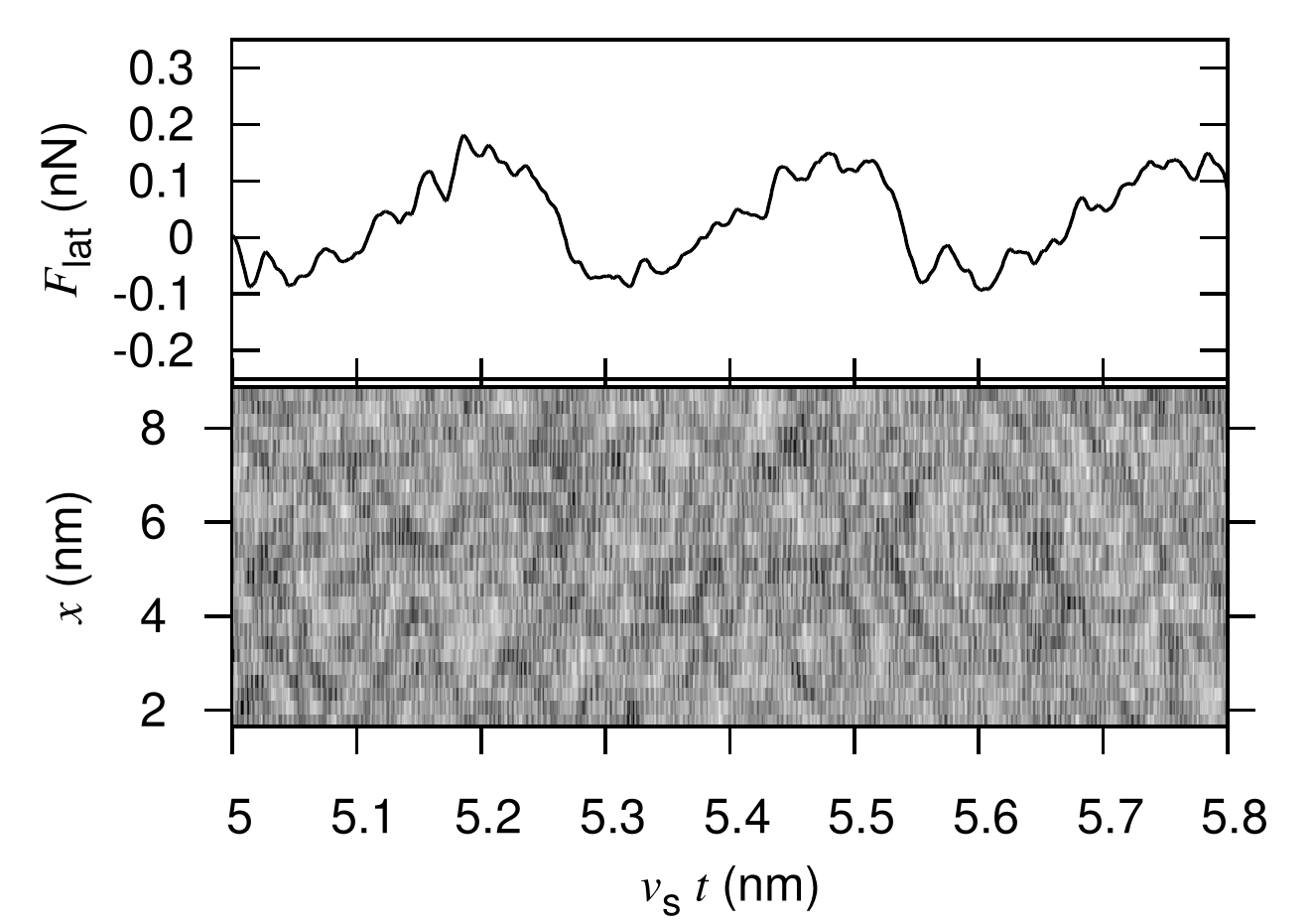,angle=270,width=8.4cm}
  \caption{
  (a) The lateral force as a function of time for the case of $E_\mathrm{ext}=-5$~V/nm (the case of Fig.~\ref{fig:chemical} of the main article).
  (b) A gray-scale map of the kinetic energy in the dipole rotation as a function of time and position of the dipole for the same simulation as in (a).
  Dark regions indicate high kinetic energy.
  During a slip energy is released, and dipole rotation propagate away from the tip, transporting and dissipating energy.
  (c) The lateral force as a function of time for the case of $E_\mathrm{ext}=0$~V/nm.
  (d) A gray-scale map of the kinetic energy in the dipole rotation as a function of time and position of the dipole for the same simulation as in (c).
  In contrast to (b), there are no propagating waves emanating from a slip point in (c).
  For this field strength, the dipole rotation does not contribute much to the dissipation.
  \label{fig:extramap}
  }
  \end{figure}
  
  Figure~\ref{fig:extramap} is for the same system as Fig.~\ref{fig:chemical} of the main article.
  It shows the lateral force as well as gray scale maps of the kinetic energy for two different field strengths, one close to the transition, where the dipole rotation contributes to the dissipation, and one away from the transition, where it does not.
  The lateral force shows the typical stick slip behavior with superimposed smaller oscillations.
  After each slip, energy is transported through the chain and away from the tip by propagating waves.
  Movies {\tt M1.avi} and {\tt M2.avi} show the same simulations, corresponding to Figs.~\ref{fig:extramap}(a,b) and (c,d) respectively.
  The mechanism of dissipation through rapid dipole rotation and rotation wave propagation can be observed in {\tt M1.avi} and Fig.~\ref{fig:extramap}b, but not in {\tt M2.avi} and Fig.~\ref{fig:extramap}d.
  The striking propagating rotation waves, which transport energy away from the tip when $E_\mathrm{ext}=-5$~V/nm, are missing for other field strengths.

\end{document}